# Observations of the delayed-choice quantum eraser using coherent photons

Sangbae Kim and Byoung S. Ham*

Center for Photon Information Processing, and School of Electrical Engineering and Computer Science, Gwangju Institute of Science and Technology, 123 Chumdangwagi-ro, Buk-gu, Gwangju 61005, South Korea
(Submitted on March 8, 2023; *bham@gist.ac.kr)

**Abstract:** Quantum superposition is the cornerstone of quantum mechanics, where interference fringes originate in the self-interference of a single photon via indistinguishable photon characteristics. Wheeler's delayed-choice experiments have been extensively studied for the wave-particle duality over the last several decades to understand the complementarity theory of quantum mechanics. The heart of the delayed-choice quantum eraser is in the mutually exclusive quantum feature violating the cause-effect relation. Here, we experimentally demonstrate the quantum eraser using coherent photon pairs by the delayed choice of a polarizer placed out of the interferometer. Coherence solutions of the observed quantum eraser are derived from a typical Mach-Zehnder interferometer, where the violation of the cause-effect relation is due to selective measurements of basis choice.

## 1. Introduction

The delayed-choice experiments proposed by Wheeler in 1978 [1] for the complementarity theory [2] have been intensively studied over the last several decades [3-16]. Although the original concept of the complementarity theory is for the exclusive nature between non-commutable entities such as position and momentum, the delayed choice experiments have been developed for the measurement control of the wave-particle duality in an interferometric system [3]. The wave-particle duality of a single photon shows a trade-off relation between the wave nature-based fringe visibility and particle nature-based which-way information [4]. The delayed choice experiments have been broadly demonstrated using thermal lights [5], entangled photons [6-8], atoms [9-11], neutrons [3], attenuated lasers [4,12,13], and antibunched single photons [14,15]. In the delayed choice, a post-control of measurements results in a paradoxical phenomenon of violation of the cause-effect relation [16]. The quantum eraser is based on the post-choice of measurements, choosing [17] or erasing [18] one of the natures. Recently, the quantum eraser has been developed for reversing a given nature via post-measurements using entangled photons [19], coherent photons [20,21], thermal lights [22], and antibunched photons [23,24].

    In the present paper, the delayed-choice quantum eraser was experimentally demonstrated using coherent photons via polarization basis controls, where the coherent photons are obtained from an attenuated continuous wave (cw) laser. Like some delayed-choice schemes [14,18,19,21], the present one is for the post-control of the pre-determined photon nature. Here, our Mach-Zehnder interferometer (MZI) composed of a polarizing beam splitter (PBS) and a beam splitter (BS) is set for the particle nature according to the Fresnel-Arago law [25] or noninteracting quantum operators [26]. Thus, the which-way information of a single photon inside the MZI is a pre-determined fact, resulting in no interference fringes in the output ports of the MZI. Without controlling the MZI itself, however, we experimentally retrieve the wave nature of the photon by controlling the output photon's polarization basis using a polarizer [14,19,21]. If the post-measurements show an interference fringe, it represents the violation of the cause-effect relation because the choice of the polarizer satisfies the space-like separation. For this, we measured first- and second-order intensity correlations using a coincidence counting unit.

## 2. Experimental setup

Figure 1 shows schematic of the present delayed-choice quantum eraser using coherent photons generated from an attenuated cw laser. For Fig. 1, a coincidence counting unit (CCU, DE2; Altera) is used for both first- and second-order intensity correlations between two detectors D1 and D2 (SPCM-AQRH-15, Excelitas). For the second-order correlation, only doubly bunched photons are counted by CCU. The generation ratio of doubly-bunched photons to single photons is ~1 % at the mean photon number $\langle n \rangle \sim 0.01$ (see Section A of the Supplemental Materials). For the first-order intensity correlation, both input channels of CCU from D1 or D2 are measured individually for a period of one second. The higher-order bunched photons are neglected by Poisson statistics (see Section A of the Supplemental Materials). To provide polarization randomness of a single photon, a 22.5°-rotated half-wave plate (HWP) is placed just before the MZI. By the followed PBS, the single photon inside the MZI shows distinguishable photon characteristics with perfect which-path information: $|\psi\rangle_{MZI} = \frac{1}{\sqrt{2}}(|V\rangle_{UP} + |H\rangle_{LP})$. Thus, the measured photons outside the MZI show the predetermined particle nature of a single photon (not shown), as in refs. [14,19,21].

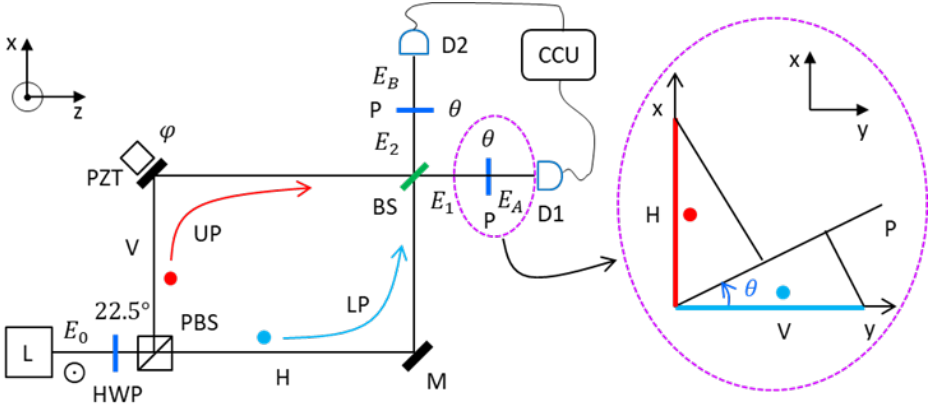

**Figure 1.** Schematic of the quantum eraser. (Dotted circle) Projection onto a polarizer. L: laser, HWP: half-wave plate, PBS: polarizing beam splitter, H (V): horizontal (vertical) polarization, M: mirror, PZT: piezo-electric transducer, BS: beam splitter, P: polarizer, D1/D2: single photon detector. CCU: coincidence counting unit. The light of laser L is vertically polarized with respect to the plane of incidence. Each colored dot indicates a single photon having the same probability amplitude.

Due to the predetermined distinguishable photon characteristics of the particle nature, the MZI does not result a φ-dependent interference fringe for the output photons ($E_1; E_2$). This is due to noninterfering quantum operators [26], as demonstrated in refs. [14,19]. Due to the classical physics of the cause-effect relation, the action of Ps outside the MZI for the output photons ($E_1; E_2$) should not change the predetermined photon nature inside the MZI. To satisfy the space-like separation, the length of each arm of the MZI is set to be 2 m, corresponding > 6 ns in the delayed choice of P. Regarding the temporal resolution (< 1 ns) of the single photon detector as well as the CCU (6 ns), the condition of the space-like separation is satisfied. Thus, any violating measurements should belong to the quantum mystery of the delayed-choice quantum eraser.

The polarizer's rotation angle θ is with respect to the vertical axis $\hat{y}$, as shown in the Inset. $E_0$ denotes an amplitude of a single photon. The mean photon number is set at $\langle n \rangle \sim 0.01$ to satisfy incoherent and independent conditions of statistical measurements, resulting in the mean photon-to-photon separation (600 m) far greater than the coherence length (3 mm) of the cw laser (see Section A of the Supplemental Materials). Doubly-bunched photon pairs are also satisfied for this condition. Thus, the measurements of Fig. 1 are for a statistical ensemble of single photons controlled by Ps.

For the MZI phase control φ, the path-length difference (ΔL) is adjusted to be far less than the coherence length $l_c$ (3 mm). This MZI coherence condition is easily tested for the same polarization-based MZI interference. Thus, the MZI in Fig. 1 satisfies a general scheme of single-photon (noninterfering) interferometers [27]. Each output photon ($E_1$ or $E_2$) from the MZI can be represented by a superposition state of the orthonormal polarization bases at equal probability amplitudes: $|\psi\rangle_{out} = \frac{1}{\sqrt{2}}(|V\rangle e^{i\varphi} + |H\rangle)$. This polarization-basis randomness of the MZI output photons is originated in the random polarization bases provided by the 22.5°-rotated HWP. In ref. [15], the measurement control with Ps in Fig. 1 is replaced by a linear optics-combined electro-optic modulator (EOM) system. By this EOM switching module, the same MZI scheme as in Fig. 1 is satisfied for the post-control of output photons [15]. Classical photon cases have also been discussed for the same results of the quantum eraser [20], where different analyses have been separately presented [22-24].

## 3. Analysis

To coherently interpret the delayed-choice quantum eraser in Fig. 1, the MZI is analyzed using a coherence approach for the output photons:

$$\begin{bmatrix} E_1 \\ E_2 \end{bmatrix} = \frac{E_0}{2}[BS][\Phi][PBS][HWP]\begin{bmatrix} 1 \\ 0 \end{bmatrix}$$

$$= \frac{E_0}{2}\begin{bmatrix} i(H + Ve^{i\varphi}) \\ H - Ve^{i\varphi} \end{bmatrix}, \quad (1)$$

where $[BS] = \frac{1}{\sqrt{2}}\begin{bmatrix} 1 & i \\ i & 1 \end{bmatrix}$, $[\Phi] = \begin{bmatrix} 1 & 0 \\ 0 & e^{i\varphi} \end{bmatrix}$, $[PBS][HWP] = \begin{bmatrix} H & iV \\ iV & H \end{bmatrix}$, and $E_0$ is the amplitude of a single photon. Here, V (H) represents a unit vector of a vertically (horizontally) polarized photon. Most importantly, interference between the H- and V-polarized photons on the BS of the MZI shows independent photon characteristics in both output ports due to orthogonal polarizations. Thus, the calculated mean intensities of $E_1$ and $E_2$ in Eq. (1) are $\langle I_1 \rangle = \langle I_2 \rangle = \langle I_0 \rangle/2$, regardless of $\varphi$. These are the coherence solutions of the MZI for the particle nature of a single photon with perfect which-way information.

By inserting a polarizer (P) outside the MZI, Eq. (1) is coherently rewritten for the polarization projection on P (see Inset of Fig. 1):

$$E_A = \frac{iE_0}{2}(sin\theta + cos\theta e^{i\varphi}), \quad (2)$$

$$E_B = \frac{E_0}{2}(sin\theta - cos\theta e^{i\varphi}), \quad (3)$$

where $\theta$ is the rotation angle of P. Thus, Eqs. (2) and (3) represent polarization projections of the output photon onto the polarizers: $\hat{V} \rightarrow \hat{p}cos\theta$ and $\hat{H} \rightarrow \hat{p}sin\theta$. Here, the positive $\theta$ is for the clockwise direction from the vertical axis of the photon propagation direction (z) (see the Inset of Fig. 1). For the negative rotation, however, the projections are denoted by $\hat{V} \rightarrow \hat{p}cos\theta$ and $\hat{H} \rightarrow -\hat{p}sin\theta$. The projection onto the polarizer P represents the action of the delayed-choice for the quantum eraser.

The calculated mean intensities of Eqs. (2) and (3) are as follows:

$$\langle I_A \rangle = \frac{\langle I_0 \rangle}{4}\langle 1 + sin2\theta cos\varphi \rangle, \quad (4)$$

$$\langle I_B \rangle = \frac{\langle I_0 \rangle}{4}\langle 1 - sin2\theta cos\varphi \rangle, \quad (5)$$

Equations (4) and (5) are the analytical solutions of the quantum eraser in Fig. 1 (see also Fig. 2). Here, the MZI coherence is for each single photon, resulting in the self-interference in the MZI [26]. Due to the low mean photon number, no coherence exists between consecutive photons, satisfying the condition of a statistical ensemble. For $\varphi = 0$, total intensity through Ps is $\langle I_A \rangle + \langle I_B \rangle = \langle I_0 \rangle/2$, regardless of $\theta$, resulting in a 50 % event loss. This selective measurement by P at the cost of 50 % event loss is the origin of the quantum eraser, as differently argued for no choice of quantum eraser [28].

For $\theta = \pm\frac{\pi}{4}$ ($\pm 45°$), Eqs. (4) and (5) are rewritten for the first-order intensity correlation:

$$\langle I_A \rangle = \frac{\langle I_0 \rangle}{4}\langle 1 \pm cos\varphi \rangle, \quad (6)$$

$$\langle I_B \rangle = \frac{\langle I_0 \rangle}{4} \langle 1 \mp \cos\varphi \rangle. \tag{7}$$

For Eqs. (6) and (7), the same P-projected photon measurements have been demonstrated in refs. [14,15] for single photons and a polarizer in ref. [19] for entangled photons. Although the EOM block control looks a direct control of the MZI [14], it actually corresponds to the combination of PBS and P in Fig. 1 (see Section B of the Supplemental Materials). In SPDC processes, entangled photons automatically satisfy both ± signs in Eqs. (6) and (7) via spatial mixing of the signal and idler photons [29]. This is the fundamental difference between coherent photons and entangled photon pairs for the quantum eraser [30]. The sum of the polarization bases in Eqs. (6) and (7), thus, corresponds to the entangled photon-pair case, as long as it deals with the first-order intensity correlation [19]. Regarding the causality violation, thus, Eqs. (6) and (7) witness the quantum feature of the delayed-choice quantum eraser for Fig. 1.

The second-order intensity correlation $R_{AB}$ via coincidence detection between D1 and D2 in Fig. 1 shows the intensity product between Eqs. (6) and (7):

$$R_{AB} = \frac{I_0^2}{4}(1 - \cos 2\varphi), \tag{8}$$

where a doubly-bunched photon pair relates to $2I_0$. Compared with ref. [19] based on entangled photons, the doubled oscillation in Eq. (8) is due to the out-of-phase fringes in D1 and D2, resulting in a classical nature. Unlike coincidence detection-caused nonlocal correlation, Eq. (8) is not for the quantum feature of a joint-phase relation [30]. This is because there is no such joint-phase action by polarizers (discussed elsewhere) [31].

### 3. Experimental results

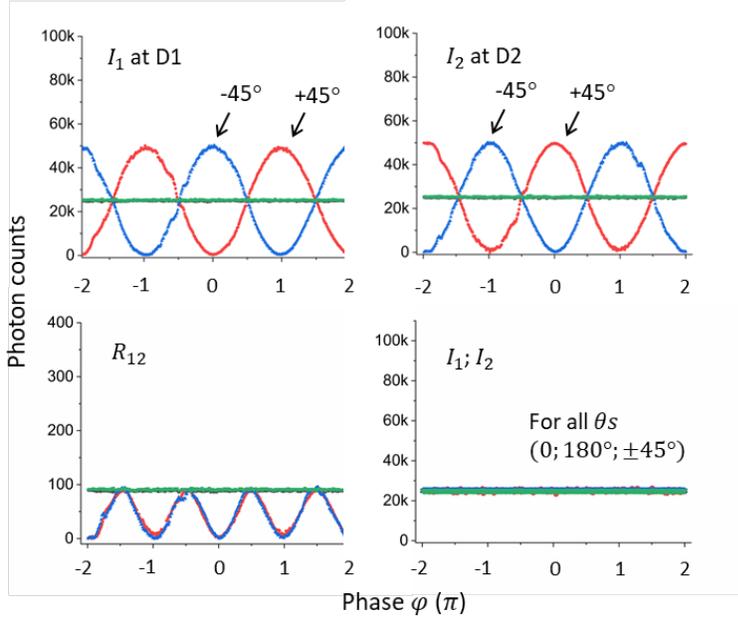

**Figure 2.** Experimental observations of the delayed-choice quantum eraser. (upper panels) Red: $\theta = 45°$, Blue: $\theta = -45°$, Green: $\theta = 0°$, Black: $\theta = 180°$. $\Delta L \ll l_c$, where $\Delta L$ is path-length difference between UP and LP. $l_c$ is coherence length of the laser L. (lower lest panel) Coincidence detection for the upper panels (color matched). (lower right panel) $\Delta L \gg l_c$ for upper panels ($\theta = \pm 45°; 0°; 180°$). Photon counts are for 0.1 s. Total data points for each $\theta$ in each panel are 360. The measured statistical error in each data is less than 1 % (see the Supplementary Materials).

The upper panels of Fig. 2 show the experimental proofs of the delayed-choice quantum eraser in Fig. 1 for coherent single photons measured by D1 and D2, respectively, for two different θs. As expected from Eqs. (6) and (7), fringes appear in both measurements for $\theta = \pm 45°$. However, no fringe appears for $\theta = 0°; 180°$, as expected by Eqs. (4) and (5) (see the overlapped green and black lines). The observed fringes represent the wave nature of the photon inside the MZI in Fig. 1. The statistical error (standard deviation) in single photon measurements is less than 1 % (see Section A of the Supplemental Materials). This is a big benefit of using coherent photons from a stabilized laser compared to entangled photons from spontaneous parametric down conversion process (SPDC) or anti-bunched photons from N-V color centers, whose respective photon counts are less than 10% [19] and 1% [14] of Fig. 2. Because the PB-MZI is not actively stabilized, most errors are from the air turbulence affecting MZI path lengths.

The lower left panel of Fig. 2 is for coincidence detection for the upper panels (color matched). The photon counts for the coincidence detection in the lower left panel is less than 1 % of those in the upper left panel of single photons. This is due to Poisson statistics for $\langle n \rangle \sim 0.01$. As expected in Eq. (8) for the coherence product, the doubled fringe oscillation period is the direct result of the intensity product between them showing the classical nature. Obviously, this intensity product of the lower left panel has nothing to do with the nonlocal quantum feature due to different purposes without independent local control parameters [19,31].

The lower right panel of Fig. 2 is for the incoherence condition of each photon by setting the MZI path-length difference ($\Delta L$) far greater than the coherence length $l_c$ of the laser. As shown, the single photon's coherence in the MZI is the key to the quantum eraser. This fact has never been discussed seriously so far, even though it seems to be obvious [16]. The observed fringes in Fig. 2 for the first-order intensity correlation demonstrate the same mysterious quantum eraser [14], because the predetermined particle nature of the photon inside the MZI (see the green line) cannot be controlled or changed by the post-measurements of the output photons [13,14,19]. Due to the benefit of coherence optics, the observed visibilities in the upper panels of Fig. 2 are near perfect.

## 5. Conclusion

The delayed-choice experiments were conducted for the quantum eraser via post-control of polarization basis of coherent photons in a coincidence detection scheme for the first-order intensity correlation. Corresponding coherence solutions were also derived in the same setups for the quantum eraser. Like conventional delayed-choice quantum eraser using orthogonal polarization bases, predetermined photon characteristics of the particle nature were retrospectively converted into the wave nature via post-selected polarization-basis projection, resulting in the violation of the cause-effect in classical physics, where the predetermined which-way information of a photons was completely erased by the post-choice of the polarizer satisfying the space-like separation. The cost of the post-measurements by the polarizer is a 50 % loss of measurement events. As usual in nonlocal quantum features, the observed quantum eraser was also due to the selective measurements of the mixed polarization bases.

**Methods**

In Fig. 1, the laser L is SDL-532-500T (Shanghai Dream Laser), whose center wavelength and coherence length are 532 nm and 3 mm, respectively. The laser light is vertically polarized. For the random but orthogonal polarizations of a single photon, a half-wave plate (HWP) is rotated by 22.5 degrees from its fast axis. For a single photon, the laser L is attenuated by neutral density filters, satisfying Poisson distribution (see the Supplementary Materials). The measurements for both output photons from the MZI are conducted by CCU (DE2; Altera) via a set of single photon detectors D1 and D2 (SPCM-AQRH-15, Excelitas). The dead time and dark count rate of the single photon detectors are 22 ns and 50 counts/s, respectively. The resolving time of the single photon detector is ~350 ps, whose converted electrical pulse duration is ~6 ns. For the polarization projection by Ps in Figs. 1, 3, and 4, four different rotation angles are set (-45, 0, 45 or 90 degrees) to the clockwise direction with respect to the vertical axis of the light propagation direction. The photon counts in Figs. 2 and 4 are measured by CCU for 0.1 s and calculated by a home-made Labview program.

In Fig. 2, the mean photon number is set at $\langle n \rangle \sim 0.01$. The maximum number of measured single photons in each MZI output port is ~a half million per second, resulting in the mean photon-to-photon distance of 600 m.

Compared with the laser's coherence length of 3 mm, it is clear that the measured single photons are completely independent and incoherent among them. On behalf of the polarizing beam splitter (PBS), perpendicularly and horizontally polarized components of an incident photon are separated into the upper (UP) and lower paths (LP), respectively. Both split components of a single photon are recombined in the BS, resulting in PB (PBS-BS)-MZI. Thus, the photons in the PB-MZI in Fig. 1 behave as the particle nature, resulting in no interference fringes in the output ports. Thus, the photons inside the MZI represent perfect which-way information or distinguishable characteristics.

The length of each arm of the PB-MZI is set at 2 m, and the path-length difference between UP and LP is kept to be far less than 3 mm to satisfy the coherence condition of each photon. This coherence condition is essential for the delayed-choice quantum eraser experiments. The φ phase control of the PB-MZI is conducted by a piezo-electric optic mount (PZT; KC1-PZ, Thorlabs) connected by a PZT controller (MDT693A, Thorlabs) and a function generator (AFG3021, Tektronix). For Figs. 2 and 5, the data is measured under the φ scanning mode, where the phase resolution is $\frac{2\pi}{180}$ radians. Thus, Figs. 2 and 5 have 180 data points for a 2π cycle of φ (see Table 1 of the Supplementary materials). The BS position for recombination of two split components of a single photon is well adjusted for a complete overlap between them.

**Author Contributions:** S.K conceived the idea, conducted experiments and provided the data. B.S.H. developed the idea, analyzed the data, and wrote the manuscript.
**Funding:** This work was supported by the ICT R&D program of MSIT/IITP (No. 2023-2022-2021-0-01810) via Development of Elemental Technologies for ultra-secure Quantum Internet. BSH also acknowledges that this work was also supported by GIST-GRI 2023.

**Conflicts of Interest:** The author declares no conflict of interest.
**Data Availability Statement:** All results and data obtained can be found in open access publications.